\documentclass[twocolumn,prl,aps,amsmath,amssymb,floatfix,showpacs]{revtex4}
\usepackage{graphicx}
\usepackage{bm}

\begin{document}
\title{Measuring the intrinsic charge transfer gap
using \\ K-edge X-ray absorption spectroscopy}

\author{C. Gougoussis, M. Calandra, A. Seitsonen, Ch. Brouder, A. Shukla and F. Mauri}
\affiliation{CNRS and 
Institut de Min\'eralogie et de Physique des Milieux condens\'es, 
case 115, 4 place Jussieu, 75252, Paris cedex 05, France}
\date{\today}

\begin{abstract}

Pre-edge features in X-ray absorption spectroscopy contain key information
about the lowest excited states and thus on the most interesting physical 
properties of the system. In transition metal oxides they are particularly
structured but extracting physical parameters by comparison with a calculation
is not easy due to several computational challenges.
By combining core-hole attraction
and correlation effects in first principles approach,
we calculate Ni K-edge X-ray absorption spectra in NiO. 
We obtain a striking, parameter-free agreement with experimental data and show 
that dipolar pre-edge features above the correlation gap are due to non-local excitations
largely unaffected by the core-hole. We show that in charge transfer insulators,
this property can be used to measure the correlation gap and probe the intrinsic position of the upper-Hubbard band.
\end{abstract}
\pacs{ }

\maketitle

% introduction

The description of electronic excitations in correlated materials is a challenge 
since many relevant phenomena, such as magnetism in transition metal (TM) compounds or
high T$_c$ superconductivity in doped Mott-insulators
are a consequence of strong electron-electron interaction.
Recently, core-hole spectroscopy unveiled 
unexpected electronic excitations in correlated antiferromagnetic insulators
\cite{Hasan, Hayashi} as well as in high T$_c$ superconductors
\cite{Collart, Kim_Hill_PRL}. 
In La$_2$CuO$_4$, K$\alpha$ resonant inelastic
X-ray scattering (RIXS) and theoretical calculations\cite{Shukla} 
demonstrated that dipolar pre-edge features 
just above the correlation gap 
are due to intersite Cu 4p-3d hybridization.
Since in correlated insulators the TM empty d-states 
form the upper Hubbard bands, it
is crucial to see (i) if 
off-site pre-edge excitations
occur in other correlated compounds \cite{Vanko}, 
(ii) to what extent their energy depends on the
presence of a core-hole in the final state. 

The presence of a core-hole substantially complicates the interpretation of
X-ray absorption (XAS) spectra. 
Since core-hole attraction shifts the empty d-states much more
than the empty p-states, typically by several eV, 
it is unclear to what extent the 
energy of the pre-edge features can provide useful information on the position 
of the corresponding excitations in the material in the absence of a core-hole.
Furthermore, in correlated insulators core-hole attraction and Hubbard repulsion 
partially compensate each other. Thus the use of fitting parameters to
describe these interactions, as is commonly done in the literature
\cite{Kotani_RMP}, does not allow to distinguish between the two effects.
A treatment of core-hole attraction and Hubbard-repulsion
from first principles is needed.

In the case of K-edge XAS spectra,
theoretical calculations are difficult since excitations 
in a large energy window above the Fermi level need to be described.
Thus methods dealing successfully with correlation effects in NiO
in an energy window close to the Fermi level (such as 
DMFT \cite{KunesPRL07} or cluster calculations \cite{TaguchiPRL08})
cannot be used to describe pre-edge and near-edge structures in K-edge
XAS spectra.
This is relevant since (i) low energy pre-edge features
provide relevant information on the local environment of the absorbing atom
\cite{Uozumi92, Joly99, Uozumi,Shukla,Vanko} and (ii) 
it has been suggested that in correlated oxides
``shakedown'' excitations occur as
near-edge structures \cite{Tolentino}.

%Several methods have been developed to describe XAS spectra, however they
%all have shortcomings in the case of correlated systems.
%Cluster calculations \cite{Kotani_RMP} are based on
%exact diagonalisation of a manybody hamiltonian with many
%empirical parameters 
%which are tuned to describe the experimental spectra.
%In the case of K-edge XAS, this method is limited to the 
%pre-edge region and inclusion of intersite effects is not evident.
%Multiple scattering \cite{Natoli} and Density functional theory (DFT) based
%methods including core-hole effects 
%\cite{taillefumier} lead to a very satisfactory description of
%K-edge XAS spectra in weakly correlated systems. However, as it will
%be shown in this work, the description of 
%correlated materials is more difficult, particularly for what concerns the 
%pre-edge features.
%Finally, the solution of the Bethe-Salpeter \cite{Shirley} equation is a
%reliable method to deal with intermediate correlation, however its applicability
%to K-edge XAS in Mott insulator is unclear and its computational 
%cost is considerable.

In this work we present a first-principles approach to describe 
the K-edge XAS spectra in correlated insulators. The method is based on a recently
developed DFT+U method and includes core-hole attraction\cite{taillefumier}. 
Differently from other 
DFT+U schemes \cite{Anisimov}, the U parameter is obtained by
linear-response \cite{cococcioni1,cococcioni2}. Consequently, {\it U is
not a fitting parameter} but an intrinsic linear response property since it measures
the spurious curvature of the energy functional as a function of occupation
\cite{cococcioni1}. 
We apply the method to the K-edge XAS of NiO,  which is the prototype 
antiferromagnetic correlated insulator and whose ground state and excitations 
are incorrectly described by 
standard DFT approaches. We demonstrate that  the pre-edge dipolar features
above the correlation gap are due to non-local excitations to second 
nearest-neighboring Ni atoms reflecting the superexchange interaction.
We show that, due to its non-local nature, the dipolar pre-edge feature
is unshifted by core-hole attraction and it is, thus, a measure of the 
upper-Hubbard band in the absence of a core-hole in the final state.
Finally we exploit the insensitivity of the pre-edge dipolar features
on core-hole attraction to show how to measure 
the charge transfer gap using K-edge XAS.

We use the NiO experimental crystal structure. The paramagnetic-cell group-space 
is then Fm\=3m, Ni occupies the 4a position and O the 4b.
The cubic lattice parameter is  $a=4.1788 \AA$ for NiO \cite{srivastava}.  
DFT calculations are performed using the 
Quantum-ESPRESSO code \cite{PWSCF}. 
We use Troullier-Martins \cite{Troullier} pseudopotentials, the spin-polarized 
generalized gradient approximation (GGA) \cite{PBE} and the recently developed
DFT+U method of ref. \cite{cococcioni1,cococcioni2}.
The wave functions are expanded using a 140 Ry energy cutoff 
for NiO. The calculated value of U for NiO 
is  U=7.6 eV.

XAS spectra are computed in a supercell approach
including core-hole effects in the pseudopotential of the absorbing atom.
From the self-consistent charge density the XAS spectra are obtained 
using the continued fraction \cite{taillefumier} and the PAW method of 
ref. \cite{Blochl}.
We considered $2\times 2\times 2$ supercells of the magnetic cell  
(32 atoms). 
We used  $4\times4\times4$ k-points grid 
both for the charge density and the continued fraction
calculation. 

\begin{figure}[t]
  \includegraphics[width=0.8\columnwidth]{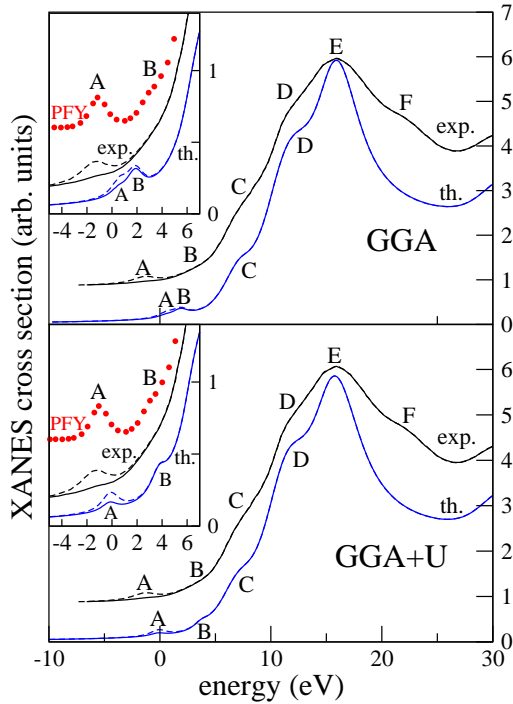}
\caption{(Color online) Calculated and measured \cite{vedrinskii}
Ni K-edge XAS of NiO including partial fluorescence 
yield (PFY) (this work). 
The dashed (solid) curves correspond to $e_g$ ($t_{2g}$) orientation.}
\label{fig:fig_nio_1}
\end{figure}

In fig. \ref{fig:fig_nio_1} we show calculated NiO XAS spectra and their
decomposition into dipolar and quadrupolar parts using
GGA and GGA+U, as compared to
experimental data \cite{vedrinskii} 
for single crystals. We consider 
the following two sets of polarization $ {\bm \epsilon}$ and wavevector ${\bf k}$:
(i)  ${\bm \epsilon} \parallel [1 1 0]$  and ${\bf k} \parallel [-1 1 0]$ ,
denoted by $e_g$ orientation (dashed lines) and 
(ii) ${\bm \epsilon} \parallel [1 0 0]$  and 
${\bf k} \parallel [0 1 0]$,
denoted by $t_{2g}$ orientation (solid lines).
The directions are in terms of the paramagnetic cubic crystal cell.
We note that U has no effect on high-energy near-edge 
and far-edge features (labeled C,D,E).
On the contrary the energy and angular dependence of 
pre-edge structures (see insets in Fig. \ref{fig:fig_nio_1})
are incorrect in the framework of GGA.
Peak B is at too low energy and a too large mixing occurs between
structures A and B.
In the GGA+U calculation the  
dipolar peak B is shifted to $\approx 2$ eV higher energies and
the angular dependence is in excellent agreement with 
experimental data. The CGA+U calculation also
shows that a very small dipolar component is present in peak A,
commonly interpreted as purely quadrupolar.
Detailed analysis of the angular dependence of the experimental NiO XAS spectra 
\cite{vedrinskii} shows that even for the $t_{2g}$ orientation,
where no quadrupolar transition occurs because the t$_{2g}$ states are occupied, 
a small peak is present \cite{ApeakNiO}, in agreement with our findings. 
Thus, in correlated insulators, the use of the DFT+U approximation
is mandatory to obtain a correct description of the pre-edge features.

\begin{figure}[t]
\includegraphics[width=0.9\columnwidth]{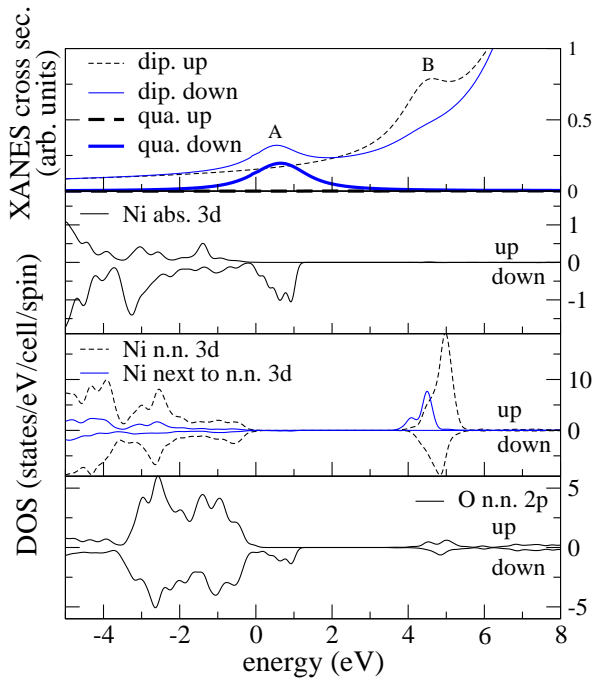}
\caption{(Color online) Comparison between GGA+U calculated K-edge Ni
XAS and L\"owdin projected density of states.}
\label{fig:fig_nio_2}
\end{figure}

To complete the understanding of the pre-edge features in NiO we 
resolve the XAS spectrum in its spin dependence and  
calculate the density of states projected over atomic orbitals using
L\"owdin projections. Without loss of generality we have considered
the absorbing Ni to be spin-up polarized. 
As can be seen in Fig. \ref{fig:fig_nio_2}, the quadrupolar part of
peak A is mostly due to intra-site excitations to d-states 
lowered by core-hole attraction. 
However, since the hybridization between Ni and O is very strong, core-hole attraction
lowers a small portion of the O 2p states generating a small
dipolar component in peak A. In a atomic orbital picture, peak A is
due to direct dipole transitions from Ni 1s states to O 2p states.
In our calculation the intensity of this excitations is of the order of 
$1\%$ of the edge jump. In experiments it is somewhat smaller because of the 
larger linewidth. This estimate is crucial for the quantitative
description of this off-site excitation in multiplet calculation.

The spin-resolved dipolar spectrum shows that
peak B is mostly due to transition to up spin-polarized states.
Since the absorbing atom has 5 up electrons in d-states,
the B excitation {\it must} have an off-site component.
L\"owdin projections demonstrate that it is due to transition 
to on-site Ni empty 4p-states hybridized to empty 3d-states of  
next-to-nearest neighbours Ni atoms. This off-site
excitation is then a fingerprint of the hopping process leading
to superexchange in NiO and
a direct probe of the upper Hubbard band in NiO.

To understand the role of core-hole effects on the A and B
features we have performed calculations with and without a core
hole in the final state. We found that  
peak A is shifted by 4 eV with respect to the edge by core-hole
attraction. On the contrary peak B is essentially unshifted. 
The difference in the behavior of the two peaks
is due to the non-local nature of peak B.

\begin{figure}[t]
\includegraphics[width=0.9\columnwidth]{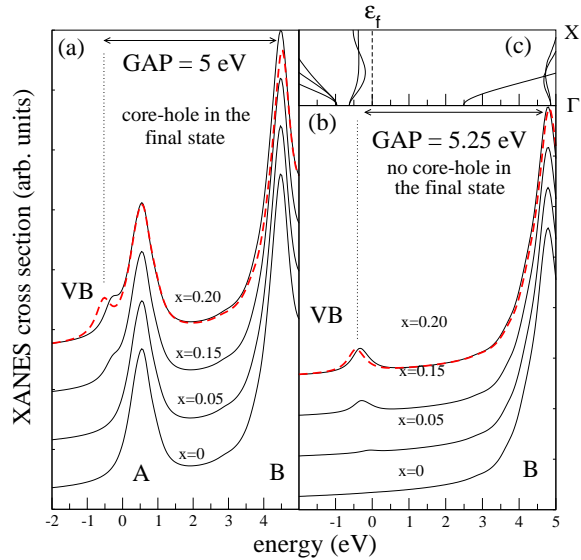}
\caption{(Color online) Calculated pre-edge features of Ni K-edge XAS 
spectra of Li$_x$Ni$_{1-x}$O with (a) and without (b) core-hole effects
 using a rigid doping (solid line) approach 
or adding a compensating charged background (red-dashed lines).
Peaks labeling is the same as in Fig. \ref{fig:fig_nio_1}, 
VB stands for Valence Band. (c) Band structure of NiO along $\Gamma$X.}
\label{fig:fig_doped}
\end{figure}

We now demonstrate that the non-local nature of the B excitation and its 
weak dependence on
core-hole effects can be used to measure the correlation gap between the
$|3d^8 \underline{L}\rangle$ and the $|3d^9 L \rangle$ states.
We consider hole-doped NiO. In practice this can be achieved through
Li doping \cite{ReinertZPHYSB1995}, as in the case of Li$_x$Ni$_{1-x}$O. 
A large range of doping can be experimentally obtained with $0 < x < 0.7$.
It has been shown \cite{KuiperPRL89,vanElp} that, 
when NiO is doped with holes, the holes mainly reside on O atoms, due to the 
charge transfer character of NiO. 
Thus the top of the NiO valence band has mainly p character.
This suggests that in Ni K-edge XAS of Li$_x$Ni$_{1-x}$O, in the low doping regime,
an additional dipolar peak should occur in the pre-edge, resulting from the
holes entering the top of the valence band. This is confirmed by
oxygen K-edge absorption 
where the top of the valence band is seen even at dopings as low as
$x=0.05$)\cite{KuiperPRL89}.
At somewhat larger doping, a similar peak due to Oxygen p-holes 
should also occur in the dipolar part of Ni K-edge XAS. 
Since no experimental data are available in the literature,
we directly simulate Ni K-edge XAS in Li$_x$Ni$_{1-x}$O.

We calculate hole-doping of NiO in two
different ways, namely (i) by rigid-band doping of NiO and (ii) adding a compensating
charge background and recalculating self-consistently the charge-density and the
XAS spectra. The results of the Ni K-edge XAS in Li$_x$Ni$_{1-x}$O in the 
pre-edge region with and without a core-hole in the final state
are shown in Fig. \ref{fig:fig_doped}. 
We find that the 
 top of the valence band should be visible in Ni K-edge XAS
at dopings of $x\approx 15-20\%$. 
In the calculated spectra, with or without a core-hole in the final state, 
the energy position of
the top of the valence band and of peak B are independent on doping.
The agreement between the rigid band picture and the 
compensating-charge background calculation validates the rigid band
picture for $0 < x < 0.2$ 

In NiO the experimentally measurable gap is due to the excitation between the top of the
valence band and the empty d-states. In DFT+U \cite{Bengone} and GW \cite{Faleev} 
calculations the
lowest single particle excitation is between the valence band and
an s band occurring at energies lower than the empty d states
(see Fig. \ref{fig:fig_doped}). 
This band is invisible in experiments since
its optical matrix elements are extremely weak \cite{Bengone}. 
Moreover its also invisible in K-edge XAS
since it is highly dispersive and has mostly Ni s and O s components.
Thus in the absence of core-hole effects, the distance between the top of the valence
band and peak B is a measure of the correlation gap, being the dependence of the
energy position on doping very weak (see Fig. \ref{fig:fig_doped}, right panel).  
When core-hole attraction is considered (see Fig. \ref{fig:fig_doped}, left panel),
we find that, despite the occurrence of a core-hole exciton (peak A),
the distance between the top of the valence band and
peak B is very weakly affected ($\approx 5\%$). 
Consequently in lightly hole-doped NiO the distance between the top
of the valence band and peak B is a measure of the correlation gap even 
in the presence of a core-hole in the final state.

In our calculation the charge-transfer gap is 5 eV
(peak-to-peak distance). 
In NiO optical absorption starts at $3.1$ eV and reaches its
maximum at $4.0$ eV \cite{Sawatzky84,HuefnerReview}. 
This value is reduced respect to our due
to excitonic effects.
In photoemission and inverse photoemission \cite{Huefner86} the gap 
(peak-to-peak distance) is roughly 5.5 eV, in fairly good agreement with our
value. Thus our proposal allows
for an independent estimate of the optical gap in NiO.

The procedure outlined for NiO can be used to measure the charge-transfer
gap and the upper Hubbard band in other charge-transfer 
insulators. In these systems the top of the valence band is due to Oxygen
p-states and it is thus visible in dipolar TM K-edge XAS
upon doping. 
If the doping necessary to detect this feature 
is low enough not to affect substantially
the electronic structure then a rigid-band doping picture applies and the
top of the valence band is at the same energy as in the undoped system.
Moreover,  hole states being dipolar in nature, the effects of core-hole
attraction are weak. The top of the valence band of the doped system
can then be used as a reference energy for the measurement of the excitations
seen in TM K-edge XAS.
The second necessary condition is the occurrence of
non-local dipolar features  \cite{Shukla,Vanko} (peak B) 
in the pre-edge region due to transitions to  d-states of neighboring TM atoms
promoted by hybridization with the TM absorbing atom p-states.
If these excitations are visible,
we have shown that they represent the upper Hubbard band of the material
and the energy difference between this excitation and the top of the 
valence band in the weakly hole-doped system is a measure of the charge transfer
gap.
Thus, TM K-edge XAS in weakly hole-doped charge-transfer insulators 
is unbiased experimental tool to measure the charge-transfer gap.

In this work we have developed a new first-principles parameter-free 
method to calculate K-edge XAS spectra including core-hole effects
in the final state and electronic correlation at the DFT+U level.
We have shown that the method provides spectra in excellent agreement 
with experimental data for NiO, the prototype correlated insulator.
We have interpreted all the pre-edge and near-edge features in the 
experimental data. In particular, we have identified a dipolar 
pre-edge feature as due to non-local excitations to d-states of the
second nearest-neighboring Ni atoms, namely the upper Hubbard band.
We have shown that, due to its non-local nature, 
this excitation is unaffected by the
presence of a core-hole in the final state.
Starting from this result we have proposed a new way to measure the 
intrinsic correlation gap in charge-transfer insulators based on TM K-edge
XAS. This method is complementary to optical measurements and 
more straightforward than a
combined photoemission-inverse photoemission experiment.

We acknowledge useful conversation with  A. Kotani, G. Dr\"ager, O.\v{S}ipr ,
Ph. Sainctavit, M. Arrio, D. Cabaret, A. Juhin, H. Kulik, N. Marzari and
L. Reining. Calculations
were performed at the IDRIS supercomputing center (project 071202).

%Unused bibitems

%


\begin{thebibliography}{99}
\bibitem{ApeakNiO} In fig. \ref{fig:fig_nio_1} of this work and in fig. 3 of 
ref. \cite{vedrinskii} a very small peak is 
present for t$_{2g}$ orientation.
\end{thebibliography}
\end{document}